\documentclass[nofootinbib,showpacs]{revtex4}

\usepackage{bm}

\begin{document}

\title{Tri-bimaximal mixing, discrete family symmetries, \\and a conjecture connecting the quark and lepton mixing matrices}

\author{Catherine I Low}
\email{c.low@physics.unimelb.edu.au}
\author{Raymond R Volkas}
\email{r.volkas@physics.unimelb.edu.au}
\affiliation{School of Physics, Research Centre for High Energy Physics, The University of Melbourne, Victoria 3010, Australia}
\date{\today}

\begin{abstract}
Neutrino oscillation experiments (excluding the LSND experiment) suggest a 
tri-bimaximal form for the lepton mixing matrix.
This form indicates that the mixing matrix is probably independent of the 
lepton masses,
and suggests the action of an underlying discrete family symmetry.
Using these hints, we conjecture that the contrasting forms of the quark and
lepton mixing matrices may both be generated by such a discrete family 
symmetry. This idea is that the diagonalisation matrices out of which the 
physical mixing matrices are composed have large mixing angles, which cancel 
out due to a symmetry when the CKM matrix is computed, but do not do so in the MNS case.
However, in the cases where the Higgs bosons are singlets under the symmetry, 
and the family symmetry commutes with $SU(2)_L$, we prove a no-go
theorem: no discrete unbroken family symmetry can produce the required mixing
patterns. We then suggest avenues for future research.
\end{abstract}

\pacs{12.15.Ff, 14.60.Pq}

\maketitle

\section{Introduction}

Experimental observations of neutrino oscillations \footnote{For the purpose 
of this study, we have assumed the LSND results 
\cite{lsnd} have a non-oscillation explanation. The reader should be aware, 
however, that this assumption may be false.} 
point to a mixing matrix of the form 
\begin{equation}\label{tbm}
U_{\mathrm{MNS}}=\left(\begin{array}{ccc}
\sqrt{\frac{2}{3}} & \frac{1}{\sqrt{3}} & 0 \\
-\frac{1}{\sqrt{6}} & \frac{1}{\sqrt{3}} & \frac{1}{\sqrt{2}} \\
-\frac{1}{\sqrt{6}} & \frac{1}{\sqrt{3}} & -\frac{1}{\sqrt{2}} \\
\end{array}
\right),
\end{equation}
where the flavour eigenstates are related to the mass
eigenstates via $(\nu_e, \nu_\mu, \nu_\tau)^T = U_{\mathrm{MNS}}
(\nu_1, \nu_2, \nu_3)^T$. Such a mixing pattern
has been termed ``tri-bimaximal mixing'' \cite{hps1}. (Majorana phases
 have not been included in the above mixing matrix as they do
not lead to observable effects in oscillations).
A mixing matrix of this form was first investigated by Wolfenstein in 
1978 \cite{wolf} 
(with degenerate mass eigenstates $\nu_1$ and $\nu_3$), and proposed 
more recently in the light of the new experimental observations by
Harrison, Perkins and Scott \cite{hps1, hs2, hs3} and He and Zee 
\cite{zeehe1, hezee2}.
The generation of small deviations from tri-bimaximal 
mixing has been investigated by Xing \cite{xing}.

The tri-bimaximal form is a very special case of the general mixing matrix
parameterised in the usual way by the Euler angles $\theta_{ij}$ where 
$i,j = 1,2,3$. 
The angle $\theta_{23}$, extracted from atmospheric neutrino
experiments \cite{Kamatm, Soudanatm, MACROatm, IMBatm}, takes the best fit value 
of $\sin^2\theta_{23}=\frac{1}{2}$ \cite{fog}. Solar neutrino results 
\cite{KamLANDsol, SNOsol, K2Ksol, SKsol, SAGEsol, HOMESTAKEsol, GNOsol}
are accommodated in Eq.\ (\ref{tbm}) through the choice 
$\sin^2\theta_{12}=\frac{1}{3}$, which is in the middle of the allowed 
``large mixing angle regions'' denoted LMA-I 
and LMA-II \cite{fogKamLand}. The third mixing angle, measured by the 
non-observation of $\nu_e$ disappearance \cite{CHOOZ}, is taken as the 
current best fit 
$\theta_{13}=0$ \cite{fog}. Note that $\theta_{23}$ takes the 
maximum possible value, while
$\theta_{13}$ takes the minimum possible value.

\subsection{Mathematics suggested by tri-bimaximal mixing}

If these special mixing angle values are indeed the correct ones, then it
is unlikely that they arise from a random choice of parameters \cite{anarchy}.
This encourages one to look for exact or approximate symmetries
of nature, operative even at low energy scales, that enforce the special
tri-bimaximal form (or something close to it).

\subsubsection{Mixing angles independent of masses}

The elements of the tri-bimaximal mixing matrix are square roots of fractions,
whereas the charged lepton masses appear to have no precise fractional
relationships, and neither do the preferred neutrino $\Delta m^2$
parameters. This motivates the construction of models where the
mixing angles, though precisely defined, are independent of the
mass eigenvalues. Such an approach is to be contrasted with the
often considered alternative proposal that relates mixing angles
 to mass ratios \cite{wein, fri3, Har, Frog}.

\subsubsection{Abelian symmetries}

Harrison, Perkins and Scott \cite{hps1} proposed weak basis mass matrices for 
charged leptons and neutrinos that generate tri-bimaximal mixing.  
An attractive feature of the proposed mass matrices is that they can be 
generated by discrete Abelian symmetries acting on the three generations of 
charged leptons and neutrinos. These symmetries dictate the form of the mixing 
matrix, but leave the masses as free parameters (see above discussion). 
The utility of these mass
matrices suggests that Abelian generation symmetries are interesting 
candidates 
for the new symmetries that might explain the neutrino mixing pattern.

\subsection{Aims of this paper}

\subsubsection{Quark and lepton mixing matrices derived from a symmetry}

In Sec.\ \ref{symsect} the Harrison, Perkins and Scott proposal 
\cite{hps1} will be reviewed. We will then extend their ideas by
conjecturing that the underlying symmetries might simultaneously
produce a quark mixing matrix that is almost the identity matrix and a
leptonic analogue that has the very different tri-bimaximal
form. While we find this an attractive hypothesis, it is not
so easy to implement in a completely well-defined extension
of the standard model. As we shall see, this proposal requires that
left-handed charged leptons and left-handed neutrinos
transform differently. But to have the symmetry group $G_{\mathrm{SM}}$
of the standard model extended to $G_{\mathrm{SM}} \otimes G_H$, where $G_H$
is a discrete horizontal or generation symmetry,
the left-handed charged leptons and left-handed 
neutrinos must transform in the same way under the symmetry, as they are 
members of the same $SU(2)_L$ doublet. 

\subsubsection{Form-diagonalisable matrices}

Section \ref{form} will define a class of matrices that are invariant under a 
symmetry and where the unitary matrices that diagonalise them 
are independent of the eigenvalues. We dub matrices such as these 
``form-diagonalisable'' and propose them as good  
candidates for lepton mass matrices because they 
generate mixing angles that are 
independent of the eigenvalues. 
This section will look at some interesting mathematics 
that relates the symmetry group to the diagonalisation matrices. 

\subsubsection{No-go theorem}

Motivated by the symmetries proposed by Harrison, Perkins and Scott 
\cite{hps1}, Sec.\ \ref{no-go} will investigate the possibility of using 
such symmetries to extend the standard model. We assume left-handed neutrinos 
transform under the symmetry in the same way as left-handed charged leptons, 
and that the Higgs bosons are singlets.
Given these assumptions, we find
that tri-bimaximal mixing, or any other form that is both
phenomenologically acceptable and predictive, 
cannot be generated by an unbroken family symmetry.

\subsubsection{Further symmetries to investigate}

Ways around the no-go theorem will be briefly discussed in Sec.\ \ref{future}.
Either or both of the
assumptions of the theorem -- that the Higgs fields are singlets and that the
symmetry is unbroken -- must be relaxed.
The generation symmetry can be extended to the Higgs sector by introducing a 
number of generations of Higgs fields that transform under 
the symmetry. Majorana neutrinos have different couplings to the Higgs 
fields from the Dirac charged leptons. As a result a symmetry that transforms
Higgs fields could potentially explain the 
differences between the mixing matrices of the leptons and the quarks.   
Vacuum expectation values of the Higgs fields can break the symmetry and 
result in different mixing matrices from those of the exact symmetry cases. 
Work along these lines is in progress. For some recent efforts, see for 
instance \cite{kubo, grimus, maa4, vallea4, maplato}.

\section{Discrete symmetries constrain mixing matrices} \label{symsect}

Many theories have been constructed using symmetries to generate
preferred mass patterns
and mixing angles. For example, democratic mass matrices can be generated 
from an $S_{3L} \times S_{3R}$ generation symmetry \cite{fri1,fri2,fuku}, 
$L_e-L_{\mu}-L_{\tau}$ symmetry leads to bimaximal mixing 
(disfavoured by the current data) \cite{moha1,lav1,lav2}, an $S_2$ permutation 
symmetry acting on $\nu_{\mu}$ 
and $\nu_{\tau}$ results in maximal atmospheric mixing \cite{Ma, mutau,lavs2}.

\subsection{How symmetries constrain mixing matrices}

The mixing matrix is related to the charged lepton mass matrix $M_\ell$ and 
the neutrino mass matrix $M_{\nu}$ in any 
weak basis by the unitary diagonalisation matrices 
$U_{\ell_L}$ and $U_{\nu}$. We use
\begin{equation}
\mathrm{Diag}(m_e,m_{\mu},m_{\tau})
=U_{\ell_L}^{\dagger}M_\ell U_{l_R},
\qquad 
\mathrm{Diag}(m_1,m_2,m_3 )=U_{\nu}^{\dagger}M_{\nu}U_{\nu}^*,
\end{equation}
to extract the lepton mixing matrix via 
\begin{equation}
U_{\mathrm{MNS}}=U_{\ell_L}^{\dagger}U_{\nu}.
\end{equation}

The symmetries of the standard model do not dictate the form of the mass 
matrices. The charged lepton mass matrix $M_\ell$ can be any $3 \times 3$ matrix, and if neutrinos are Majorana, then $M_{\nu}$ must be symmetric, but
is otherwise unconstrained.  As a result the mixing matrix can be of 
any unitary form, and 
the masses are unrestricted by the standard model symmetries.
However, if a generation symmetry holds, the form of the mass matrices 
-- and hence the mixing matrix -- are constrained. 
For the Lagrangian to be invariant under transformations of 
the three generations of Majorana neutrinos, the left-handed charged leptons 
and the right handed charged leptons, 
\begin{equation}
\nu \rightarrow X_{\nu} \nu, \qquad  \ell_L \rightarrow X_L \ell_L, \qquad  \ell_R \rightarrow X_R \ell_R,
\end{equation}
the mass matrices must obey the restrictions, 
\begin{equation}\label{symdefn}
M_{\nu}=X_{\nu}^{\dagger} M_{\nu} X_{\nu}^*, 
\qquad   M_\ell=X_L^{\dagger} M_\ell X_R,
\end{equation}
where $X_{\nu}$, $X_L$ and $X_R$ are  $3 \times 3$ unitary matrices.
The special case of the vector-like symmetry would have left and right-handed 
fields transforming 
identically, with $X_L=X_R$. 

\subsection{Harrison, Perkins and Scott's proposed symmetries}\label{hpssect}

Harrison, Perkins and Scott \cite{hps1} suggested mass matrices of the form
\begin{equation} \label{hpsmm}
M_{\ell}=\left(\begin{array}{ccc}
a & b & c\\
c & a & b\\
b & c & a
\end{array}\right), \qquad  
M_{\nu}=\left(\begin{array}{ccc}
x & 0 & y\\
0 & z & 0\\
y & 0 & x
\end{array}\right),
\end{equation}
where the parameters $a,b,c$ are related to the three charged lepton masses, 
and $x,y,z$ provide three independent neutrino masses. 
The charged lepton mass matrix is of circulant form and can be generated by a 
cyclic permutation  ($C_3$) 
symmetry. An $S_2 \times S_2$ symmetry generates the neutrino mass matrix. 

The unitary transformation matrices are 
\begin{equation} \label{hpssym}
X_{L 1}=X_{R 1}=
\left(\begin{array}{ccc} 0 & 0 & 1 \\ 1 & 0 & 0 \\ 0 & 1 & 0 
\end{array}\right);
\qquad   X_{L 2}=X_{R 2}=
\left(\begin{array}{ccc} 0 & 1 & 0 \\ 0 & 0 & 1 \\ 1 & 0 & 0 
\end{array}\right);
\qquad X_{\nu 1} =\left(\begin{array}{ccc} 0 & 0 & 1 \\ 0 & 1 & 0 \\ 1 & 0 & 0  \end{array}\right); 
\qquad 
X_{\nu 2} =
\left(\begin{array}{ccc} 1 & 0 & 0 \\ 0 & -1 & 0 \\ 0 & 0 & 1  \end{array}\right). 
\end{equation}

The proposed mass matrices are diagonalised by
\begin{equation}\label{hpsu}
U_{\ell_L}=U_{\ell_R}=\frac{1}{\sqrt{3}}\left(\begin{array}{ccc}
1 & 1 & 1 \\
1 & \omega & \omega^* \\
1 & \omega^* & \omega
\end{array}\right),
\qquad
U_{\nu}=
\left(\begin{array}{ccc}
\frac{1}{\sqrt{2}} & 0 & -\frac{1}{\sqrt{2}} \\
0 & 1 & 0 \\
\frac{1}{\sqrt{2}} & 0 & \frac{1}{\sqrt{2}} \end{array} \right),
\end{equation}
where $\omega \equiv e^{2 \pi i/3}$,
which combine to give tri-bimaximal mixing. 

\subsection{Using the symmetry to constrain quark mixing to small angles: a 
conjecture}\label{conjecture}

Harrison, Perkins and Scott's idea can be extended to include the quarks, and 
produce small quark mixing.
We conjecture that the up-type quarks and the down-type quarks transform
under the $C_3$ generation symmetry in the same way as the charged leptons 
transform above. This will force both quark mass matrices into circulant form
\begin{equation}
M_{u}=\left(\begin{array}{ccc}
a_u & b_u & c_u\\
c_u & a_u & b_u\\
b_u & c_u & a_u
\end{array}\right), \qquad  
M_{d}=\left(\begin{array}{ccc}
a_d & b_d & c_d\\
c_d & a_d & b_d\\
b_d & c_d & a_d
\end{array}\right).
\end{equation}
These mass matrices are diagonalised by the same matrix $U_u=U_d$,
resulting in 
$U_{\mathrm{CKM}}=U_u^{\dagger} U_d=I$, corresponding to no quark mixing. 
As with the leptons, all quark masses are unrestricted by the symmetry.

The unbroken symmetry produces $U_{\mathrm{CKM}}=I$, and 
$U_{\mathrm{MNS}}$ to be of tri-bimaximal form. Small symmetry 
breaking can be introduced to generate off-diagonal terms in the
quark mixing matrix. This breaking may also deviate the lepton mixing 
matrix away from tri-bimaximal form.

The quarks transform together, whereas the
neutrinos transform independently of the charged leptons. This accounts for the
 differences between the quark and the lepton mixing matrices.

Under the symmetry the neutrinos transform in a different way from 
all the other fermions. This  may be associated with other special 
characteristics of the neutrinos, for example,
the Majorana nature of the neutrino, or the lack of electric charge.

\subsection{$SU(2)_L$ constraint on standard model extensions}

The conjecture outlined above shows that discrete generation symmetries 
can produce tri-bimaximal lepton mixing and 
small quark mixing.
However, these symmetries cannot be incorporated into an extension of the
standard model with the structure $SU(2)_L \otimes G_H$, where the  $G_H$ is the discrete horizontal or family symmetry. The symmetries of 
Eq.\ (\ref{hpssym}) do not commute with $SU(2)_L$, as the left-handed neutrinos
transform under the symmetry in a different way from the left-handed charged 
leptons, whereas a symmetry that is an extension to the standard model should 
preserve the standard model symmetry $SU(2)_L$, by having members of the same 
$SU(2)_L$ doublet
transform together \cite{moha}. $SU(2)_L$ is not violated by the quark transformations as the up and down-type quarks transform in the same way.

This constraint makes it difficult to find any symmetry that gives rise to
tri-bimaximal mixing.
Sec.\ \ref{no-go} investigates whether it is possible for any discrete family
 symmetry
to predict tri-bimaximal mixing when the $SU(2)_L$ constraint is included.

\section{Form-Diagonalisable Matrices}\label{form}

\subsection{Definition}
A form-diagonalisable matrix is a matrix that is invariant under a symmetry, 
and with diagonalisation matrices whose elements depend on the form of the 
original matrix only. 
As a result the diagonalisation matrices are independent of the matrices' 
eigenvalues. 

An $n \times n$ form-diagonalisable matrix is defined by
\begin{equation}\label{formdefn}
F=\sum_i^k \alpha_i \lambda_i
\end{equation}
where: 
\begin{itemize}

\item $\lambda_i$ are $n \times n$ matrices of pure numbers, and $\alpha_i$ are $n$ complex parameters;

\item $\lambda_i$ are simultaneously diagonalisable by two unitary matrices 
$U_L$ and $U_R$, where $U_L^{\dagger}\lambda_i U_R$ is diagonal for all $i$;

\item $\lambda_i$ are invariant under a group transformation: $\lambda_i=X_L^{\dagger}\lambda_i X_R$;

\item $k \leq n$. 

\end{itemize}
Note that for $k<n$, only $k$ eigenvalues are independent.

These conditions result in the masses being linear combinations of $\alpha_i$,
 and the diagonalisation matrices, $U_L$ and $U_R$, being independent of 
these masses.

\subsection{Examples of form-diagonalisable matrices with Abelian symmetries}

Equation (\ref{hpsmm}) has two examples of form diagonalisable mass matrices, 
with the symmetries being the Abelian groups $C_3$ and $S_2 \times S_2$. 

The form of the mass matrices is dependent not only on the symmetry group, but
also on the representation of the group that the transformation matrices $X_L$ and $X_R$ take.

\subsubsection{Regular representation of Abelian groups}

An interesting relationship occurs between the symmetry group and the 
diagonalisation matrix when the symmetry is an Abelian group in the regular 
representation.
The regular representation of a group of order $n$, is a set of $n$  matrices
$X_i$. The matrices are unitary, have size $n \times n$, and their elements 
are $0$ or $1$. A matrix $M$ is considered to be invariant under the regular 
representation of a group when 
$M= X_i^T M X_i$ for all $i$. 

For Abelian symmetries the 
mass matrix that is invariant under the regular representation is a linear 
combination of all the representation matrices themselves, i.e. $\lambda_i$ 
of Eq.\ (\ref{formdefn}) are the $X_i$.  This is shown in App.\ \ref{regreps}.

The matrix $U$ that diagonalises the mass matrix $M$ can be simply derived 
from the $n$ one-dimensional representations of the group $G$: Each column of 
the diagonalisation matrix is made up of a normalised list of the elements of 
the one-dimensional representations, and each column corresponds to a 
different one-dimensional representation.
As all the irreducible representations of Abelian groups are one dimensional, 
the character table lists these representations, and the diagonalisation 
matrix can be read directly off the table.

\subsubsection{$C_3$ example}

This relationship between the regular representation and the diagonalisation 
 matrices is illustrated by the $C_3$ symmetry of the charged leptons outlined
 in Sec.\ \ref{hpssect}.  

The charged lepton mass matrix of Eq.\ (\ref{hpsmm}) is invariant under the 
regular representation of $C_3$ which is given by
\begin{equation}
\Bigg\{\left(\begin{array}{ccc}
1 & 0 & 0\\
0 & 1 & 0 \\
0 & 0 & 1 \\
\end{array}
\right) , \left(\begin{array}{ccc}
0 & 1 & 0\\
0 & 0 & 1 \\
1 & 0 & 0 \\
\end{array}
\right) , \left(\begin{array}{ccc}
0 & 0 & 1\\
1 & 0 & 0 \\
0 & 1 & 0 \\
\end{array}
\right) \Bigg\}.
\end{equation}
The mass matrix is made up of a linear combination of invariant matrices 
$\lambda_i$. In this case the $\lambda_i$ are the 
representation matrices themselves, forming the mass matrix $M_\ell$ of Eq.\ (\ref{hpsmm}).
 The diagonalisation matrix is 
\begin{equation}\label{ul}
U_{\ell}=\frac{1}{\sqrt{3}}\left(\begin{array}{ccc}
1 & 1 & 1 \\
1 & \omega & \omega^* \\
1 & \omega^* & \omega
\end{array}\right)
\end{equation}
where $\omega=e^{2 \pi i/3}$, $ \omega^*$, $ 1$ 
are the cube roots of unity. This diagonalisation matrix can be constructed 
using  the one-dimensional representations of $C_3$ which are  
$\{1,1,1\}$, $\{1,\omega, \omega^* \}$, $\{1,\omega^*, \omega \}$. 
Each column of the diagonalisation matrix is made up of a one-dimensional 
representation, and the matrix is normalised.

Representations other than the regular representations can also produce 
form-diagonalisable mass matrices. An example of this is the $S_2 \times S_2$ 
symmetry which generates the mass matrix $M_{\nu}$ of
Eq.\ (\ref{hpsmm}). In cases other than the regular representation,
the relationship between 
the representation of the symmetry and the diagonalisation matrix is not clear.

\section{No-go Theorem for Discrete Family symmetries}\label{no-go}
 
Individual lepton 
number symmetry $U(1)_{L_e}\otimes U(1)_{L_\mu} \otimes U(1)_{L_\tau}$
is a symmetry of the 
standard model with massless neutrinos, and is known to be broken
by neutrino oscillations. However, if a discrete subgroup of this 
symmetry is unbroken by the neutrino mass term, this will constrain the form of
the mixing matrix.

The success of the symmetries in Eq.\ (\ref{hpssym}) in generating 
tri-bimaximal mixing, and the idea that a subgroup of 
$U(1)_{L_e}\otimes U(1)_{L_\mu} \otimes U(1)_{L_\tau}$ may still remain 
unbroken with massive neutrinos motivates the systematic study of discrete 
Abelian group symmetries, with the added 
constraint of having the left-handed charged leptons transform in the same
way as the left-handed neutrinos.

This section shows that discrete unbroken generation symmetries (Abelian and non-Abelian) with the $SU(2)_L$ constraint and the other assumptions stated below
cannot generate tri-bimaximal mixing. In fact, the only mixing matrix that 
falls within experimental bounds and is generated by a symmetry, is 
the mixing matrix that is completely unrestricted by the symmetry.
In this section we assume that the Higgs 
bosons are singlets of the symmetry.

Section \ref{no-gononab} shows that discrete non-Abelian generation symmetries
give rise to degenerate charged leptons, proving that non-Abelian symmetries 
cannot produce mass and mixing schemes that agree with experiment.

Section \ref{lconstrain} considers how Abelian groups can constrain the 
charged lepton Dirac mass matrix. Exactly how the transformations alter the 
neutrino mass matrix depends on 
the type of mass term, because Majorana mass terms are constrained by the 
symmetry in
a different way from Dirac mass terms. Because of this the no-go theorem for
Abelian groups is segmented into three cases; Majorana neutrinos (Sec.\ 
\ref{no-goab}), Dirac 
neutrinos (Sec.\ \ref{no-godirac}), and Majorana neutrinos when the mass term 
is generated by the seesaw mechanism (Sec.\ \ref{no-goseesaw}).
 In the seesaw case we assume that the right-handed Majorana mass matrix is 
invertible.

We show that in all three cases all mixing schemes that can be produced by 
Abelian
symmetries are not allowed by experiment, except for the case where the
mixing is not constrained by the symmetry at all.

\subsection{Equivalent representations yield identical mixing}\label{secequiv}

The matrices $X_{L i}$ and $X_{R i}$ of 
Eq.\ (\ref{symdefn}) that transform the leptons are representations of the 
symmetry group. Different 
representations of the same symmetry group provide different restrictions on 
the mass matrices. As there are three generations of leptons we are 
interested in three dimensional representations only. A given symmetry group 
has an infinite number of three dimensional representations, but only a finite 
number of inequivalent representations.

Two different representations $X_i$ and $Y_i$, are considered to be
equivalent if they are related by a 
similarity transformation
\begin{equation}
Y_i=V^{\dagger}X_i V,
\end{equation}
where $V$ is any unitary matrix. 

Appendix \ref{equivreps} shows that two equivalent transformation matrices 
restrict the mixing matrix in an 
identical way. This is because 
the weak basis leptons $(\nu_{L_X},l_{L_X})^T$
in the case where
the representation $X_i$ is chosen, are related to the weak basis leptons $(\nu_{L_Y}, l_{L_Y})^T$ in 
the $Y_i$ case by a basis change, as per

\begin{equation}
\left(\begin{array}{c}\nu_{L_X}\\ l_{L_X}\end{array}\right) \rightarrow  \left(\begin{array}{c}\nu_{L_Y}\\ l_{L_Y}\end{array}\right)=V^\dagger \left(\begin{array}{c}\nu_{L_X}\\ l_{L_X}\end{array}\right).
\end{equation}  

Since the mixing matrix is associated with the mass basis of the leptons, not 
the weak basis, the two equivalent representations will restrict the mixing 
matrix in an identical way. 
As there are only a finite number of inequivalent representations of 
any discrete group it is possible to find
 all mixing matrices that can be generated by a given group. 

All Abelian representations are equivalent to a diagonal representation -- a 
representation where all matrices are diagonal. The converse is also true; no 
non-Abelian representation has matrices that are all diagonal 
(as diagonal matrices commute). This provides a 
convenient way of analysing many groups at once. First we will consider at 
non-Abelian groups by examining how non-diagonal transformations affect 
mass matrices and mixing, and then we consider Abelian representations by 
looking at diagonal representations.

\subsection{Non-Abelian groups}\label{no-gononab}

Non-Abelian groups have Abelian (for example the trivial representation) and 
non-Abelian representations. Abelian 
representations of non-Abelian groups are not faithful, and are also 
representations of Abelian groups.
This section shows that non-Abelian representations constrain some charged 
leptons to be degenerate. Abelian representations are covered by sections 
\ref{no-goab}, 
\ref{no-godirac} and \ref{no-goseesaw}.

As explained in Sec.\ \ref{secequiv}, two equivalent representations correspond 
to two different bases. So if the mass matrices are invariant under some
non-Abelian transformation, there exists a non-Abelian representation of the 
group that corresponds to the  charged lepton mass basis 
$M_\ell=\textrm{Diag}(m_e,m_\mu,m_\tau)$.
As this representation is non-Abelian, there 
is at least one matrix that is not diagonal.

Mass degeneracy can be concluded by considering just one non-diagonal 
transformation matrix. For example a block diagonal unitary matrix
\begin{equation}
X_L=\left(\begin{array}{ccc}x &0 &0\\0& y &w \\0&z&v \end{array}\right),
\end{equation}
constrains $M_\ell M_\ell^\dagger$ by
\begin{eqnarray}
M_\ell M_\ell^\dagger&=&X_L^\dagger M_\ell M_\ell^\dagger X_L
=\left(\begin{array}{ccc}x^* &0 &0\\0& y^* &z^* \\0&w^*&v^* \end{array}\right)
\left(\begin{array}{ccc}m_e^2 &0 &0\\0& m_\mu^2 & 0\\0&0&m_\tau^2 \end{array}\right)
\left(\begin{array}{ccc}x &0 &0\\0& y &w \\0&z&v \end{array}\right)\\
&=&\left(\begin{array}{ccc}m_e^2 |x|^2 &0 &0\\
0& m_\mu^2 |y|^2+ m_\tau^2 |z|^2& m_\mu^2 y^* w+ m_\tau^2 z^* v\\
0&m_\mu^2 y w^*+ m_\tau^2 z v^*&m_\tau^2 |v|^2+ m_\mu^2 |w|^2\end{array}\right).
\end{eqnarray}
The $2 \times 2$ block in $X_L$ rotates $m_\mu^2$ and $m_\tau^2$, so the 
diagonal mass matrix will only be invariant under this transformation if 
$ m_\mu^2 = m_\tau^2$. 
An $X_L$ that is not in block diagonal form will result in three degenerate
charged leptons.

The same argument also applies when the $X_R$ transformation is non-Abelian.
In this case the $X_R$ transformation 
constrains $ M_\ell^\dagger M_\ell=\textrm{Diag}(m_e^2,m_\mu^2,m_\tau^2)$ by
$M_\ell^\dagger M_\ell=X_R^\dagger M_\ell^\dagger M_\ell X_R$, also resulting 
in degenerate masses.

\subsection{Abelian representations and charged lepton mass matrices}\label{lconstrain}

In the case of Abelian groups, every representation is equivalent
to a diagonal matrix
representation, so to find out all the mixing matrices that can be produced 
by an Abelian group, we can restrict the study to how mass matrices can be 
constrained by diagonal representations.

The diagonal representations 
\begin{equation}
X_L=\textrm{Diag}(e^{i\phi_1}, e^{i\phi_2}, e^{i\phi_3}), \qquad
X_R=\textrm{Diag}(e^{i\sigma_1}, e^{i\sigma_2}, e^{i\sigma_3}),
\end{equation}
constrain the charged lepton mass matrix
$M_\ell$  by $M_\ell=X_L^{\dagger} M_\ell X_R$, or, more explicitly,
\begin{equation}
M_\ell=\left(\begin{array}{ccc} \label{mlres}
r & s & t \\ u & v & w \\x & y & z 
\end{array}\right)=
\left(\begin{array}{ccc}
 r e^{-i(\phi_1-\sigma_1)} & s e^{-i(\phi_1-\sigma_2)} & 
t e^{-i(\phi_1-\sigma_3)} \\ 
u e^{-i(\phi_2-\sigma_1)} & ve^{-i(\phi_2-\sigma_2)} & 
w e^{-i(\phi_2-\sigma_3)}\\
x e^{-i(\phi_3-\sigma_1)} & ye^{-i(\phi_3-\sigma_2)} & 
z e^{-i(\phi_3-\sigma_3)}
\end{array}\right).
\end{equation}

Not all of the information contained in the mass matrix is required in order 
to find the masses and the mixing matrix.
One may simply compute the hermitian squared mass matrix $M_\ell M_\ell^\dagger$ and then diagonalise it via the left-handed matrix $U_{\ell_L}$ only, as per  $U_{\ell_L}^\dagger M_\ell M_\ell^\dagger U_{\ell_L}=\textrm{Diag}(m_e^2,m_\mu^2,m_\tau^2)$. Now, $M_\ell M_\ell^{\dagger}$ is restricted by the $X_L$ transformation by 
\begin{equation}
M_\ell M_\ell^{\dagger}=\left(\begin{array}{ccc} 
a & b & c \\ 
b^* & d & f \\
c^* & f^* & g \end{array}\right) 
=X_L^\dagger M_\ell M_\ell^{\dagger} X_L=\left(\begin{array}{ccc}
a & b e^{-i(\phi_1-\phi_2)} & c e^{-i(\phi_1-\phi_3)} \\ 
b^*e^{i(\phi_1-\phi_2)} & d & f e^{-i(\phi_2-\phi_3)} \\
c^* e^{i(\phi_1-\phi_3)} & f^* e^{i(\phi_2-\phi_3)} & g\end{array}\right).
\end{equation}
The $X_L$ transformation constrains the hermitian squared mass matrix in the 
following way:
The diagonal elements of  
$M_\ell M_\ell^{\dagger}$ are unrestricted by the symmetry; when 
$\phi_i=\phi_j$ the $ij$th term in 
$M_\ell M_\ell^{\dagger}$ is unrestricted by the symmetry; 
otherwise the $ij$th element will be zero.

Note that $M_\ell M_\ell^\dagger$ can also be constrained by the $X_R$ matrix. 
For example, if $X_L=I$ and $X_R=-I$ then $M_\ell=M_\ell M_\ell^\dagger=0$, 
even 
though the $X_L$ transformation does not constrain the mass matrix.

To make the no-go theorem simpler, we look first at how 
$U_{\mathrm{MNS}}$
can be constrained by the $X_L$ transformation, before analysing how the 
$X_R$ transformation alters the situation. For nearly all choices of $X_L$,
the $X_L$ tranformation constrains $M_\ell M_\ell^\dagger$ and $M_\nu$ in such
a way to force the mixing matrix $U_\mathrm{MNS}$ into a form that has been 
ruled out experimentally. In these cases the $X_R$ transformations are 
irrelevant, the symmetry having been ruled out for all possible choices of 
$X_R$.

\subsection{Abelian representations and Majorana neutrinos}\label{no-goab}

The left-handed transformation $X_L$ restricts the Majorana neutrino mass 
matrix by
\begin{equation}
M_{\nu}=\left(\begin{array}{ccc}
A & B & C \\
B & D & E \\
C & E & F 
\end{array}\right)=
\left(\begin{array}{ccc}
Ae^{-2i\phi_1} & Be^{-i(\phi_1+\phi_2)} & C e^{-i(\phi_1+\phi_3)} \\
Be^{-i(\phi_1+\phi_2)} & De^{-2i\phi_2} & E e^{-i(\phi_2+\phi_3)} \\
Ce^{-i(\phi_1+\phi_3)} & Ee^{-i(\phi_2+\phi_3}) & F e^{-2i\phi_3}
\end{array}\right).
\end{equation}

The $X_L$ transformation multiplies each element of the mass matrix by a 
phase. If the phase equals $1$, then the element is unconstrained by the
symmetry. If the phase is not equal to $1$, then the matrix element is forced 
to be zero.  
If $e^{i \phi_i}=\pm 1$, then the $i i$th element of the matrix will be 
unrestricted by the symmetry. 
If $e^{i \phi_i}=e^{- i \phi_j}$ then the $ij$th 
element 
will be unrestricted. Otherwise the elements will be zero.

We have performed an exhaustive analysis of all possible forms of 
lepton mixing matrices 
that can be produced by an Abelian generation symmetry. The mixing matrices
 are listed below.
Interchanging columns 
corresponds to relabeling neutrino mass eigenstates.

In the following matrices $s \equiv \sin \theta$ and $c \equiv \cos \theta$, 
where $\theta$ is unconstrained by the symmetry. The phases $e^{i \delta_i}$ 
are not neccesarily physical. 

\begin{eqnarray}
\textbf{Mixing matrix}& \textbf{Form of $X_L$ required for all $X_L$}\nonumber\\
&\nonumber\\
U_{\mathrm{MNS}_1}
=\left(\begin{array}{ccc}c e^{i \delta_1} & s e^{i \delta_2} & 0\\ -s e^{i \delta_3} & c e^{i \delta_4}&0 \\ 0&0 & 1\end{array}\right)
&
\begin{array}{c}
X_L=\textrm{Diag}(e^{i \phi_1},e^{i \phi_1},\pm 1)\\
X_L=\textrm{Diag}(\pm 1,\pm 1,e^{i \phi_3})\\
X_L=\textrm{Diag}(\pm 1,\pm 1,\mp 1)
\end{array}
\\
U_{\mathrm{MNS}_2}
=\left(\begin{array}{ccc}
1 & 0&0 \\
0& c e^{i \delta_1} & s e^{i \delta_2} \\ 
0&-s e^{i \delta_3} & c e^{i \delta_4} 
\end{array}\right)
&
\begin{array}{c}
X_L=\textrm{Diag}(\pm 1,e^{i \phi_2},e^{i \phi_2})\\
X_L=\textrm{Diag}(e^{i \phi_1},\pm 1,\pm 1)\\
X_L=\textrm{Diag}(\pm 1,\mp 1,\mp 1)
\end{array}
\\
U_{\mathrm{MNS}_3}
=\left(\begin{array}{ccc}
 c e^{i \delta_1}&0 & s e^{i \delta_2} \\ 
0& 1 &0\\
-s e^{i \delta_3}& 0& c e^{i \delta_4} 
\end{array}\right)
&
\begin{array}{c}
X_L=\textrm{Diag}(e^{i \phi_1},\pm 1,e^{i \phi_1})\\
X_L=\textrm{Diag}(\pm 1,e^{i \phi_1},\pm 1)\\
X_L=\textrm{Diag}(\mp 1,\pm 1,\mp 1)
\end{array}
\\
U_{\mathrm{MNS}_4}
=\left(\begin{array}{ccc} 
-\frac{1}{\sqrt{2}} e^{i \delta_5} & \frac{1}{\sqrt{2}} e^{i \delta_5}& 0 \\
\frac{s}{\sqrt{2}} e^{i \delta_1} & \frac{s}{\sqrt{2}} e^{i \delta_1} & c e^{i \delta_2} \\
\frac{c}{\sqrt{2}} e^{i \delta_3} & \frac{c}{\sqrt{2}} e^{i \delta_3} & -s e^{i \delta_4} 
\end{array}\right)
&
X_L=\textrm{Diag}(e^{i \phi_1},e^{-i \phi_1},e^{-i \phi_1})
\\
U_{\mathrm{MNS}_5}
=\left(\begin{array}{ccc} 
\frac{s}{\sqrt{2}} e^{i \delta_1} & \frac{s}{\sqrt{2}} e^{i \delta_1} & c e^{i \delta_2} \\
\frac{c}{\sqrt{2}} e^{i \delta_3} & \frac{c}{\sqrt{2}} e^{i \delta_3} & -s e^{i \delta_4} \\
-\frac{1}{\sqrt{2}} e^{i \delta_5} & \frac{1}{\sqrt{2}} e^{i \delta_5}& 0
\end{array}\right)
&
X_L=\textrm{Diag}(e^{i \phi_1},e^{i \phi_1},e^{-i \phi_1})
\\
U_{\mathrm{MNS}_6}
=\left(\begin{array}{ccc} 
\frac{s}{\sqrt{2}} e^{i \delta_1} & \frac{s}{\sqrt{2}} e^{i \delta_1} & c e^{i \delta_2} \\
-\frac{1}{\sqrt{2}} e^{i \delta_5} & \frac{1}{\sqrt{2}} e^{i \delta_5}& 0 \\
\frac{c}{\sqrt{2}} e^{i \delta_3} & \frac{c}{\sqrt{2}} e^{i \delta_3} & -s e^{i \delta_4} 
\end{array}\right)
&
X_L=\textrm{Diag}(e^{i \phi_1},e^{-i \phi_1},e^{i \phi_1})
\\
U_{\mathrm{MNS}_7}
= \left(\begin{array}{ccc} 1/\sqrt{2} &  1/\sqrt{2} & 0 \\ -1/\sqrt{2}&  1/\sqrt{2}&0\\0 &0 & 1 \end{array} \right)
&
\begin{array}{c}
X_L=\textrm{Diag}(e^{i \phi_1},e^{-i \phi_1},e^{-i \phi_3})\\
X_L=\textrm{Diag}(e^{i \phi_1},e^{-i \phi_1},\pm 1)
\end{array}
\\
U_{\mathrm{MNS}_8}
= \left(\begin{array}{ccc} 1/\sqrt{2} &0&  1/\sqrt{2} \\0&1&0\\ -1\sqrt{2}& 0& 1/\sqrt{2} \end{array} \right)
&
\begin{array}{c}
X_L=\textrm{Diag}(e^{i \phi_1},e^{i \phi_2},e^{-i \phi_1})\\
X_L=\textrm{Diag}(e^{i \phi_1},\pm 1,e^{-i \phi_1})
\end{array}
\\
U_{\mathrm{MNS}_9}
= \left(\begin{array}{ccc} 1&0&0\\0&1/\sqrt{2} &  1/\sqrt{2}\\ 0&-1/\sqrt{2}&  1/\sqrt{2} \end{array} \right)
&
\begin{array}{c}
X_L=\textrm{Diag}(e^{i \phi_1},e^{i \phi_2},e^{-i \phi_2})\\
X_L=\textrm{Diag}(\pm 1,e^{i \phi_2},e^{-i \phi_2})
\end{array}
\\
U_{\mathrm{MNS}_{10}}
= \textrm{Trivial -- massless neutrinos}
&
\begin{array}{c}
X_L=\textrm{Diag}(e^{i \phi_1},e^{i \phi_2},e^{i \phi_3})\\
\phi_i \neq \pm 1 \textrm{ for at least one $X_L$, for 
all $i$},\\ 
\phi_j \neq \phi_i \textrm{ for at least one $X_L$, for 
all $i$, $j$.}
\end{array}
\\
U_{\mathrm{MNS}_{11}}= \textrm{Unrestricted by the symmetry}
&
X_L=\pm I
\end{eqnarray}

In cases $U_{\mathrm{MNS}_{4,5,6}}$, $m_1=-m_2$ and $m_3=0$.
In cases $U_{\mathrm{MNS}_{7,8,9}}$, the two mixed neutrinos have $m_i=-m_j$.

Except for the case where the mixing is unrestricted by the symmetry, none of 
the above mixing matrices fall within experimental bounds. 
In the unrestricted case $U_\nu$ is unrestricted, so although right-handed 
charged lepton transformations can alter $U_{\ell_L}$, the mixing matrix
$U_\mathrm{MNS}=U_{\ell_L}^\dagger U_\nu$ will remain unconstrained by 
the symmetry.

\subsection{Abelian representations and Dirac neutrinos}\label{no-godirac}

An Abelian symmetry constrains the neutrino Dirac mass matrix in the same way
as the charged lepton Dirac mass matrix , Eq.\ (\ref{mlres}), 
except that the right-handed neutrino may transform in a different way to
the right-handed charged leptons.

Dirac neutrino mass matrices are diagonalised by
Diag$(m_1,m_2,m_3)=U_{\nu_L}^\dagger M_\nu U_{\nu_R}$, and the mixing matrix
incorporates only the left diagonalisation matrices. $U_{\nu_L}$ can be 
obtained
from $M_\nu M_\nu^\dagger$ which is restricted by the $X_L$ transformation by
$M_\nu M_\nu^\dagger=X_L^\dagger M_\nu M_\nu^\dagger X_L$.

The possible $U_\mathrm{MNS}$ matrices obtainable by the left-handed 
transformation are listed below. It is possible that the right-handed 
transformations will be able to further restrict the mixing matrices.

\begin{eqnarray}
\textbf{Mixing matrix}\:\:\:
&\textbf{Form of $X_L$ required for all $X_L$}\nonumber\\
& \nonumber\\
U_{\mathrm{MNS}_1}=\left(\begin{array}{ccc}c_l & s_l &0 \\ -s_l e^{i \delta_l} & c_l e^{i \delta_l}&0 \\0 &0 & 1
\end{array}\right)
& 
X_L=\textrm{Diag}(e^{i \phi_1},e^{i \phi_1},e^{i \phi_3})
\\
U_{\mathrm{MNS}_2}=\left(\begin{array}{ccc} 1 &0 &0 \\ 
0& c_l & s_l \\ 
0& -s_l e^{i \delta_l}& c_l e^{i \delta_l}
\end{array}\right) 
& X_L=\textrm{Diag}(e^{i \phi_1},e^{i \phi_2},e^{i \phi_2})
\\
U_{\mathrm{MNS}_3}=\left(\begin{array}{ccc}c_l &0 & s_l \\ 0& 1 &0 \\-s_l e^{i \delta_l} & 0& c_l e^{i \delta_l}
\end{array}\right)
&
 X_L=\textrm{Diag}(e^{i \phi_1},e^{i \phi_2},e^{i \phi_1})
\\
U_{\mathrm{MNS_4}}= \textrm{ unrestricted by $X_L$ }
&
X_L=e^{i\phi_1} I
\\
U_{\mathrm{MNS}}=I 
&
\begin{array}{c}
\textrm{ $e^{i\phi_1}\neq e^{i\phi_2}$ for some $X_L$, }\\ 
\textrm{$e^{i\phi_1} \neq e^{i\phi_3}$ for some $X_L$,}\\
\textrm{and  $e^{i\phi_2}\neq e^{i\phi_3}$ for some $X_L$}.
\end{array}
\end{eqnarray}

The only $U_{\mathrm{MNS}}$ that fits in with experiment is the one that is 
unrestricted by $X_L$, which occurs when $X_L=e^{i \phi}I$. In this case both 
$U_{\ell_L}$ and $U_{\nu_L}$ are unconstrained by the $X_L$ transformation. 
However,  $U_{\ell_L}$ and $U_{\nu_L}$ can be restricted by the right-handed
transformations $X_{\ell_R}$ and $X_{\nu_R}$.
If one or both of the two diagonalisation matrices remains unrestricted under 
the right-handed transformations, then $U_\mathrm{MNS}=U_{\ell_L}^\dagger U_{\nu_L}$ will be unrestricted, independent of how the second diagonalisation 
matrix is restricted by the symmetry. 

The transformation
\begin{equation}
X_{\ell_R}=\textrm{Diag}(e^{i\sigma_1}, e^{i\sigma_2}, e^{i\sigma_3})
\end{equation}
restricts the charged lepton mass matrix by
\begin{eqnarray}
M_\ell&=&\left(\begin{array}{ccc}r & s & t \\ u & v & w \\x & y & z 
\end{array}\right)\\
&=&X_L^\dagger M_\ell X_{\ell_R}
= \left(\begin{array}{ccc} e^{-i (\phi-\sigma_1)}r & e^{-i (\phi-\sigma_2)}s 
& e^{-i (\phi-\sigma_3)}t \\ e^{-i (\phi-\sigma_1)}u & e^{-i (\phi-\sigma_2)}v 
& e^{-i (\phi-\sigma_3)}w \\e^{-i (\phi-\sigma_1)}x & e^{-i (\phi-\sigma_2)}y 
& e^{-i (\phi-\sigma_3)}z 
\end{array}\right).
\end{eqnarray}
Either the $i$th column is unrestricted by the symmetry, 
($\phi = \sigma_i$),
or the symmetry constrains column $i$ to be a column of zeros 
($\phi \neq \sigma_i$). A matrix that has one column of zeros has one 
massless charged lepton.
A matrix that has no columns of zeros is completely unconstrained by the 
symmetry, and will give an unrestricted $U_{\ell_L}$. 

Therefore, in the case where $X_L=e^{i \phi} I$, $U_{\mathrm{MNS}}$ is 
unrestricted unless one or more of the charged leptons are massless. As there
are no massless charged leptons, we can conclude that for Dirac neutrinos no 
mixing matrix is
compatible with experiment, except for when $U_\mathrm{MNS}$ is completely 
unconstrained by the symmetry.

In fact, if the electron is taken to be massless (corresponding to a single 
column of zeros), we are convinced that
$U_{\ell_L}$ is also completely general, and hence, the mixing matrix is 
unrestricted by the symmetry. In this case $U_{\ell_L}$ has the 
same number of free parameters as a completely unconstrained diagonalisation 
matrix. This has been backed up by numerical calculations. 
The right-handed diagonalisation matrix $U_{\ell_R}$, however, is restricted
by the right-handed transformation.

\subsection{Abelian representations and Seesaw neutrinos}\label{no-goseesaw}

Majorana neutrino mass matrices that are generated by the seesaw mechanism
can be expressed as
\begin{equation}
M_\nu=M_d^T M_M^{-1} M_d,
\end{equation}
where $M_d$ is the Dirac mass matrix, and $M_M$ is the right-handed 
Majorana mass matrix. This equation is valid when $M_M$ is invertible. 
In this section we assume that $M_M$ is invertible. (If the 
Majorana mass matrix was not invertible, and had rank $n>3$,  
the physical particles would be $n$ ultralight neutrinos, $n$ heavy neutrinos 
and $2n-6$ neutrinos whose masses are naturally the same size as the other 
fermions \cite{glasseesaw, fukuseesaw}).

Under the $X_L$ transformations $M_\nu$ is restricted by
\begin{equation}
M_\nu=X_L^\dagger M_\nu X_L^*,
\end{equation}
the same as when the neutrinos are Majorana but do not have mass terms 
generated by the seesaw mechanism. Section \ref{no-goab} lists all the ways 
that $X_L$ can restrict the mixing matrix.
Again, the only mixing matrix that fits with experiment is the mixing matrix 
that is unrestricted by the symmetry, which occurs when $X_L=\pm I$. In this 
case the diagonalisation matrices $U_{\ell_L}$ and $U_\nu$ are both 
unrestricted by the $X_L$ transformation, but can be further restricted by 
right-handed transformations.

The right-handed charged lepton transformation restricts the mass matrix by
\begin{eqnarray}
M_\ell&=&\left(\begin{array}{ccc}r & s & t \\ u & v & w \\x & y & z 
\end{array}\right)\\
&=&X_L^\dagger M_\ell X_{\ell_R}
=\pm \left(\begin{array}{ccc} e^{i \sigma_1}r & e^{i \sigma_2}s 
& e^{i\sigma_3}t \\ e^{i \sigma_1}u & e^{i \sigma_2}v 
& e^{i\sigma_3}w \\e^{i\sigma_1}x & e^{i\sigma_2}y 
& e^{i\sigma_3}z 
\end{array}\right).
\end{eqnarray}

The argument in the Dirac neutrino section is applicable here also. Either a 
column of the mass matrix is unrestricted by the symmetry, or it is zero. If 
all columns are unrestricted, $U_{\ell_L}$ is unrestricted by the symmetry, 
giving a mixing matrix that is unconstrained by the symmetry. For each column 
that is 
constrained to be zero, there is a corresponding massless charged lepton 
which is not seen in nature.
If one charged lepton is taken to be massless, the mixing is still 
unconstrained by the symmetry.
Therefore, the only mixing matrix that can be generated by a discrete unbroken
symmetry, and is consistent with experiments is the mixing matrix that is 
completely unconstrained by the symmetry.

\section{Conclusions and future work}\label{future}

It is tantalising to suppose that a family symmetry could simultaneously 
explain both the lepton and the quark 
mixing matrices. We have shown however, that given certain assumptions, 
unbroken symmetries 
acting on the generations of the fermions cannot produce a lepton mixing matrix
of tri-bimaximal form, or anything approaching this form. 
Relaxing the assumptions of this no-go theorem may make it possible for a 
symmetry to generate an experimentally allowed mixing matrix.

An option for trying to generate non-trivial mixing in the lepton sector, 
while still 
including the $SU(2)_L$ restriction, is to utilise the different mass generation 
mechanisms for the neutrinos and charged leptons.
Charged lepton masses come from Yukawa couplings with 
the standard model Higgs doublet. Majorana neutrinos will gain masses from 
another mechanism, possibly using the same Higgs doublets in the seesaw 
mechanism, or by interaction with a Higgs triplet, or by a different mechanism.
 
If the Higgs sector is extended by introducing a number of generations of Higgs
 fields, these Higgs fields can also transform under the symmetry. Since the 
action of the Higgs fields in creating mass matrices is different for 
neutrinos compared to charged leptons, different 
restrictions for the two mass matrices will in general result. 
This in turn will lead to the 
diagonalisation matrices for neutrinos being different from that of the charged
leptons, possibly resulting in phenomenologically acceptable lepton mixing. 

Since both up-like and down-like quarks are Dirac particles, 
the action of the Higgs fields in creating their mass matrices is similar 
for both sectors. It might be possible, then, to construct a model
whereby these mass matrices are sufficiently similar so as to yield
very similar left-diagonalisation matrices. The resulting $U_{\mathrm{CKM}}$
may then be approximately diagonal, in agreement with the observed form
of this matrix. This kind of setting --  
models with a non-minimal Higgs sector -- may the appropriate one in which to
realise our conjecture (see Sub-Sec.\ \ref{conjecture}) within a complete and consistent
standard model extension, despite its original inspiration coming from
the rather different Harrison, Perkins and Scott proposal.

\acknowledgments{RRV would like to thank Tony Zee and Lincoln Wolfenstein for 
useful discussions during the {\it Neutrinos: Data, Cosmos, and Planck Scale} 
workshop held at the Kavli Institute for Theoretical Physics at the University
of California -- Santa Barbara. 
This work was supported in part by the Australian Research 
Council, and in part by The University of Melbourne.}

\bibliography{discretesym}

\appendix

\section{Regular representations of Abelian groups}\label{regreps}

For a group of rank $n$, the regular representation involves $n$, 
$n \times n$ matrices, with elements $0$ and $1$. Each row or column contains 
one $1$. The $ij$th term equals $1$ for one and only one matrix in the 
representation. 
One of the matrices is the identity.

$M$ is invariant under the regular representation of an Abelian group
if $M$ commutes with all $X$:
\begin{equation}
M=X_a^T M X_a \textrm{ for all $a$}, 
\end{equation}
where $a$ labels the $X$ matrices, or for each element
\begin{equation}
M_{ij}=\sum_{kl} (X^{a T})_{ik} M_{kl} X^{a}_{lj} \textrm{ for all $a$}. 
\end{equation}

As the group is Abelian, all the $X$ matrices commute with each other,  
so an arbitrary linear combination of the $X$ matrices will also commute 
with all $X$. The following argument shows that if $M$ commutes with $X$, 
 {\it the most general} $M$ must be a linear combination of the $X$ matrices.

The restriction forces the diagonal elements of $M$ to be equal:
\begin{eqnarray}
M_{11}&=\sum_{kl}(X^{a T})_{1k} M_{kl} X^{a}_{l1}&\\
&=(X^T)_{1j}M_{jj}X_{j1}=M_{jj}& \textrm{ choosing the $X$ to be the one that has $X_{j1}=1$}\nonumber\\
&=M_{jj}.&\nonumber
\end{eqnarray}
Since there exists a matrix $X$ such that $X_{j1}=1$ for all $j$, all the 
diagonal elements are equal. The diagonal elements
of $M$ can be written as $M_{11} I$.

By looking just at the restrictions placed on the mass matrix by an $X$ that 
has $X_{ij}=1$, we show that if $X_{kl}$ also equals $1$, then the $kl$th element of the mass
matrix must be equal to the $ij$th element, $M_{ij}=M_{kl}$.

Let us take the $X$ that has $X_{12}=1$.
\begin{eqnarray}
M_{12}&=\sum_{kl}=(X^T)_{1k}M_{kl}X_{l2}& \textrm{ choose the $X$ that has $X_{12}=1$.}\\
&= \sum_{k} (X^T)_{1k}M_{k1}X_{12}& \textrm{ choose a $k$ such that $X_{k1}=1$} \nonumber\\
&= M_{k1}&\nonumber\\
M_{k1}&=\sum_j (X^T)_{kj}M_{jk}X_{k1}& \textrm{ choose $j$ such that $X_{jk}=1$}\nonumber\\
&=M_{jk},&\nonumber
\end{eqnarray}
Repeating this will show that the restrictions from the $X$ that has 
$X_{12}=1$, ensure that
$M_{12}=M_{ij}$ if $X_{ij}=1$. 
$M_{12} X$ describes the $ij$ terms of the mass matrix, where $X_{ij}=1$.

The same argument can be made for any $M$ element. 
If $X_{ij}=X_{kl}$ for a given $X$, then 
$M_{ij}=M_{kl}$, showing that the $kl$th elements of $M$ can be expressed as
$M_{ij} X$. Therefore $M$ is a linear combination of the $X$ matrices.

\section{Proof that two equivalent representations constrain the mixing matrix in an identical way}\label{equivreps}

This proof assumes that Higgs bosons are singlets of the generation symmetry, 
and that the generation symmetry commutes with $SU(2)_L$ meaning $\nu_L$ 
transforms in the same way as $\ell_L$.
The seesaw section assumes that the right-handed Majorana mass matrix is
invertible.
 
\subsection{Charged leptons}\label{applrep}

$A_{Li}$ and $B_{Li}$ are equivalent representations which will transform the  
left-handed leptons. Each matrix is labelled by an index $i$.
$A_{\ell_{R}i}$and $B_{\ell_{R}i}$ are also equivalent representations 
which transform the right-handed charged leptons:
\begin{equation}
U_1^\dagger A_{Li} U_1=B_{Li}, \qquad
U_2^\dagger A_{\ell_{R}i} U_2=B_{\ell_R i}.
\end{equation}

The two different representations restrict the charged lepton mass matrix by
\begin{eqnarray}
M_{\ell A}=A_{Li}^\dagger M_{\ell A} A_{\ell_Ri} \:\: \textrm{ for all $i$},
\qquad &
M_{\ell B}&=B_{Li}^\dagger M_{\ell B} B_{\ell_Ri}\:\: \textrm{ for all $i$}\\ 
&&=U_1^\dagger A_{Li}^\dagger U_1 M_{\ell B} U_2^\dagger A_{\ell_{R}i} U_2.\nonumber
\end{eqnarray}

$U_1 M_{\ell B} U_2^\dagger$ has the same restrictions as 
$M_{\ell A}$. As we assume that the mass matrices are completely unconstrained
apart from the generation symmetry constraints, we can set
\begin{equation}
U_1 M_{\ell B} U_2^\dagger=M_{\ell A}.
\end{equation}

$M_{\ell}$ is diagonalised by $U_{\ell_L}$ and $U_{\ell_R}$ via
\begin{equation}
\textrm{Diag}(m_e,m_\mu,m_\tau) = U_{l_L A}^\dagger M_{l A} U_{l_R A}
=U_{l_L B}^\dagger M_{l B} U_{l_R B},
\end{equation}
so $U_{\ell_L B}=U_1^\dagger U_{\ell_L A}$ and 
$U_{\ell_R B}=U_2^\dagger U_{\ell_R A}$.

\subsection{Majorana neutrinos}\label{appmrep}

The two representations restrict the neutrino mass matrix by
\begin{eqnarray}
M_{\nu A}=A_{Li}^\dagger M_{\nu A} A_{Li}^* \textrm{ for all $i$}, \qquad &
M_{\nu B}&=B_{Li}^\dagger M_{\nu B} B_{Li}^*  \textrm{ for all $i$},\\
& &= U_1^\dagger A_{Li}^\dagger U_1 M_{\nu B} U_1^\dagger A_{Li} U_1.\nonumber
\end{eqnarray}
$U_1 M_{\nu B} U_1^T$ has the same restrictions as $M_{\nu A}$, and we can equate $U_1 M_{\nu B} U_1^T=M_{\nu A}$.

$M_{\nu}$ is diagonalised by $U_{\nu}$ via
\begin{equation}
\textrm{Diag}(m_1,m_2,m_3) = U_{\nu A}^\dagger M_{\nu A} U_{\nu A}^*
=U_{\nu B}^\dagger M_{\nu B} U_{\nu B}^*
\end{equation}
So $ U_{\nu B}=U_1^\dagger U_{\nu A}$.

Combining this result with the charged lepton results we see
\begin{equation}
U_{\mathrm{MNS}B}=U_{\ell_L B}^\dagger U_{\nu B}=U_{\ell_L A}^\dagger U_1 
U_1^\dagger U_{\nu A}=U_{\ell_L A}^\dagger U_{\nu A}=U_{\mathrm{MNS A}}
\end{equation}
showing that representation $A$ gives the same mixing matrix restrictions 
as representation $B$.

\subsection{Dirac neutrinos}\label{appdiracrep}

The right-handed neutrinos transform by the representations
$A_{\nu_R i}$ and $B_{\nu_Ri}$ which are related by
\begin{equation}
U_3^\dagger A_{\nu_R i} U_3=B_{\nu_R i}.
\end{equation}
An identical argument to App.\ \ref{applrep}
shows $U_1 M_{\nu B} U_3^\dagger$ has the same restrictions as $M_{\nu A}$, 
enabling us to set $U_1 M_{\nu B} U_3^\dagger=M_{\nu A}$.
so $U_{\nu_L B}=U_1^\dagger U_{\nu_L A}$.
$U_{\nu_R B}=U_3^\dagger U_{\nu_R A}$.

Combining this with the charged lepton result we see that the mixing matrix 
for $A$ is the same as the mixing matrix for $B$:
\begin{equation}
U_{\mathrm{MNS B}}=U_{l_L B}^\dagger U_{\nu B}=U_{l_L A}^\dagger U_1 U_1^\dagger U_{\nu A}=U_{l_L A}^\dagger U_{\nu A}=U_{\mathrm{MNS A}},
\end{equation}
showing that the two equivalent representations restrict the mixing in the same
way.

\subsection{Seesaw neutrinos}
This section assumes that the Majorana mass matrix is invertible, so the resultant light neutrino mass matrix is given by
$M_\nu=M_d^T M_M^{-1} M_d$.

From App.\ \ref{appdiracrep},
$(U_1 M_{d B} U_3^\dagger)$ has the same restrictions as $M_{d A}$, so set them
to be equal.

From App.\ \ref{appmrep}, the right-handed Majorana mass term constraints show
$(U_3^* M_{M B} U_3^\dagger)$ has the same restrictions as $M_{M A}$, so they
can be set equal.

The resultant light neutrino mass term has the restrictions
\begin{eqnarray}
M_{\nu A} &=& M_{d A} M_{M A}^{-1} M_{d A}^T\\
& = &  (U_1 M_{d B} U_3^\dagger) (U_3 M_{M B}^{-1} U_3^T)(U_3^* M_{d B}^T U_1^T)\nonumber\\
&=& U_1 M_{d B} M_{M B}^{-1} M_{d B}^T U_1^T\nonumber\\
&=&U_1 M_{\nu B} U_1^T. \nonumber
\end{eqnarray}

So $M_{\nu A}$ and $M_{\nu B}$ are related by a basis change - the same as the case with non-seesaw Majorana neutrinos.

Diagonalising:
\begin{eqnarray}
\textrm{Diag}(m_1,m_2,m_3)& =& U_{\nu A}^\dagger M_{\nu A} U_{\nu A}^*\\
&=&U_{\nu A}^\dagger U_1 M_{\nu B} U_1^T U_{\nu A}^*\nonumber\\
&=&U_{\nu B}^\dagger M_{\nu B} U_{\nu B}^*\nonumber
\end{eqnarray}
So $ U_{\nu B}=U_1^\dagger U_{\nu A}$.

So the mixing matrices for the two representations are
\begin{equation}
U_{\mathrm{MNS B}}=U_{l_L B}^\dagger U_{\nu B}=U_{l_L A}^\dagger U_1 U_1^\dagger U_{\nu A}=U_{l_L A}^\dagger U_{\nu A}=U_{\mathrm{MNS A}}
\end{equation}
Therefore, two different, but equivalent, representations restrict the mixing 
matrix in the same way.

\end{document}